\documentclass{bioinfoArXiv} 
\copyrightyear{2016}
\pubyear{2016}
\usepackage{verbatim} 
\usepackage{amssymb}
\usepackage{amsmath}
\usepackage{algorithm2e}
\usepackage{verbatim} 
\usepackage{epstopdf}
\usepackage{setspace}
\usepackage{amsthm}
\usepackage{rotating}
\usepackage[T1]{fontenc}
\usepackage{subfig}
\newtheorem{definition}{Definition}[section]

\newtheorem{theorem}{Theorem}

\begin{document}
\firstpage{1}
\title[Identification of repeats in DNA sequences]{Identification of repeats in DNA sequences using nucleotide distribution uniformity}
\author[Yin, C.]{Changchuan Yin \footnote{To whom correspondence should be addressed. Email: cyin1@uic.edu}}
\address{Department of Mathematics, Statistics and Computer Science\\ The University of Illinois at Chicago, Chicago, IL 60607-7045, USA\\}

\history{} 

\editor{} 

\maketitle

\begin{abstract}
\section{Motivation:}
Repetitive elements are important in genomic structures, functions and regulations, yet effective methods in precisely identifying repetitive elements in DNA sequences are not fully accessible, and the relationship between repetitive elements and periodicities of genomes is not clearly understood. 

\section{Results:}
We present an \textit{ab initio} method to quantitatively detect repetitive elements and infer the consensus repeat pattern in repetitive elements. The method uses the measure of the distribution uniformity of nucleotides at periodic positions in DNA sequences or genomes. It can identify periodicities, consensus repeat patterns, copy numbers and perfect levels of repetitive elements. The results of using the method on different DNA sequences and genomes demonstrate efficacy and accuracy in identifying repeat patterns and periodicities. The complexity of the method is linear with respect to the lengths of the analyzed sequences.

\section{Availability:}
The Python programs in this study are freely available to the public upon request or at https://github.com/cyinbox


\section{Contact:} \href{cyin1@uic.edu}{cyin1@uic.edu}
\end{abstract}

\section{Introduction}
\label{Introduction}
Repetitive elements in DNA sequences consist two or more copies of approximate patterns of nucleotides and are abundant in both prokaryotic and eukaryotic genomes. Over two-thirds of the human genome and $5$-$10\%$ bacterial genomes are repetitive regions \citep{de2011repetitive}. Repetitive elements play important roles in genome structure and functions such as nucleoprotein complex formation, chromosome structure, and gene expression. Various diseases including cancer and neurodegentive disease can also arise from changes of repetitive elements. The distribution of repetitive DNA sequences can be used as fingerprints of bacterial genomes \citep{versalovic1991distribution} and human individuals. 

Repetitive elements are complex structures. They may exist as imperfect tandem repeats, insertion and deletions in repeats, interspersed repeats, and palindromic sequences, etc. These partial and hidden repeat signals in DNA sequences are difficult to analyze through straightforward observation and sequence comparison. 

Currently, repetitive elements and hidden periodicities of DNA and protein sequences are primarily detected by digital signal processing and statistical approaches \citep{treangen2011repetitive}. In most signal processing methods, DNA sequences are converted to numerical sequences, and the hidden periodicities arising from repetitive elements can be identified by Fourier power spectrum at specific periodicities \citep{yin2016periodic}. Commonly used signal processing methods by Fourier transform include SRF maps \citep{sharma2004spectral}, spectral analysis \citep{buchner2003detection}, Ramanujan-Fourier transform \citep{yin2015novel}, and the periodic power spectrum method \citep{yin2016periodic}. The statistical methods are based on distribution analysis of nucleotides in DNA sequences. The common statistical methods for repeat findings are tandem repeats finder \citep{benson1999tandem} and statistical spectrum \citep{epps2011statistical}, maximum likelihood estimation \citep{arora2007detection}, and information decomposition \citep{korotkov2003information}. Besides signal processing and statistical approaches, sequence alignments such as RepeatMask are also used to identify  repetitive patterns in genomes, and but require a known reference repeat sequence.

Despite significant advances in repeat finding, it is still difficult to precisely capture the essential features of repetitive elements such as consensus patterns, perfect levels and copy numbers of repeats. For example, while Fourier transform is the most common used approach for finding repeats, it may not exactly correlate the strength of Fourier power spectrum with the perfect level of repeat patterns. Furthermore, since Fourier power spectrum is weak for short DNA sequences and long harmonious periodicities are embedded in short periodicities, Fourier transform can not capture repeats in short DNA sequences and long harmonious periodicities. Moreover, the relationship between repetitive elements and periodicities of genomes is not fully understood. Thus there is a high potential for improving the accuracy for identifying repetitive elements and better understanding the relationship of periodicities and repeats in DNA sequences \citep{suvorova2014comparative,epps2011statistical,illingworth2008criteria}.

In this paper, we present an $\textit{ab initio}$ method to quantitatively identify repetitive sequences and periodicities in DNA sequences. The method is based on the nucleotide distribution uniformity at periodic positions in DNA sequences or genomes. The distribution uniformity of nucleotides reflects the unbalance of nucleotide frequencies on periodic positions and thus can indicate the strength for periodic signals in DNA sequences. The method can also reveal the consensus repeat pattern for the major periodicity of DNA sequences, and quantitatively determine the perfect level and copy numbers of repetitive sequences. The proposed method also formulates the relationship between repetitive elements and the corresponding periodicities in DNA sequences.
 
\section{Methods and Algorithms}
\subsection{Periodic nucleotide frequencies in a DNA sequence}
A DNA molecule consists of four linearly joined nucleotides, adenine (A), thymine (T), cytosine (C), and guanine (G). A DNA sequence can be represented as a string of the four characters A, T, C, and G. The nucleotide frequencies at periodic positions, which can be represented by a congruence derivative vector \citep{wang2012Some,yin2016periodic,wang2016fast}, reflect the arrangement of repetitive elements and inner periodicities in a DNA sequence. The congruent derivative vector of a nucleotide $\alpha$ for a specific periodicity is constructed by the cumulative occurring frequencies of nucleotides at the periodic positions (Definition 2.1).
  
\begin{definition} For a DNA sequence of length $n$, let $u_\alpha  (k)=1$ when the nucleotide $\alpha$ appears at position $k$, otherwise, $u_\alpha (k)=0$, where $\alpha  \in \{ A,T,C,G\}$ and $k = 1, \cdots ,n$. The congruence derivative vector of the nucleotide $\alpha$ of the sequence for periodicity $p$, is defined as\\
\begin{equation}
\begin{gathered}
  f_{\alpha ,j}  = \sum\limits_{\bmod (k,p) = j} {u_\alpha  (k)}  \hfill \\
  j = 1, \cdots ,p,k = 1, \cdots ,n \hfill \\ 
\end{gathered} 
\end{equation}
where $mod(k,p)$ is the modulo operation and returns the remainder after division of $k$ by $p$, and $f_\alpha  = \left( {f_\alpha (1),f_\alpha (2), \cdots ,f_\alpha (p)} \right)$.
\end{definition}

Four congruence derivative vectors  $f_\alpha$ of periodicity $p$ for nucleotides A, T, C and G form a congruence derivative (CD) matrix of size $4 \times p$. The columns of the CD matrix indicate nucleotide frequencies at the periodic positions $k = pt-q$, where $k$ is the position index of a DNA sequence, $t = 1, 2, \ldots$, and $q = p - 1, \ldots , 2, 1, 0$. For example, consider the CD matrix of periodicity 5 for DNA sequence, the first column of the CD matrix shows the nucleotide frequencies at periodic positions $k = 1, 6, 11, \ldots , 5t - 4$; the second column of the matrix shows the nucleotide frequencies at periodic positions $k = 2, 7, 12, \ldots , 5t - 3$; the third column of the matrix shows the nucleotide distributions at periodic positions $k = 3, 8, 13, \ldots , 5t - 2$, and so on. In this way, the CD matrix of a DNA sequence describes nucleotide frequencies at all periodic positions and can be used to efficiently compute Fourier power spectrum and determine periodicities in the DNA sequence \citep{yin2016periodic}. In this study, instead of Fourier transform, we use the CD matrix to identify repetitive elements and periodicities directly. This approach offers elaboration of the repetitive elements such as the consensus repeat pattern, copy number and perfect level. 

To illustrate the nucleotide frequencies in the CD matrix, an artificial DNA sequence of 80 bp, ATTGAACCCGGTGCGAACAC-
ATCTCCACTCATCGAAGCGCGTCGGATCAGGTCAGATCGG-
ATCTGATCGGAACGGCTGCG, which contains 5 bp approximate repeats, is constructed. The CD matrix of periodicity 5 for this sequence is shown in Fig.1. The base with the highest frequency in each column is labeled in color, while the other bases are labeled in black. The consensus sequence of the repeats can be identified as ATCGG from the highest frequency in each column in the CD matrix. Thus, we may determine that the DNA sequence contains 16 copies of approximate ATCGG (Fig.1). The bases show in black in columns are then the mismatched nucleotides when the actual sequence is compared with the consensus sequence at the periodic positions.

 \begin{figure}
 \begin{minipage}[b]{0.50\linewidth}
    \centering%
     \begin{tabular} { |l|l|l|l|l|l| }%
       \hline%
        j & 1 & 2 &  3 & 4 &  5 \\ 
       \hline%
       G & 3 & 1 &  2 & 7 &  10 \\ 
       C & 2 & 1 &  13 & 3 &  4 \\ 
       A & 11 & 3 &  0 & 3 &  2 \\ 
       T & 0 & 11 &  1 & 3 &  0 \\ \hline%
     \end{tabular}%
     \par\vspace{0pt}
   \end{minipage}
   \begin{minipage}[b]{0.50\linewidth}
        \centering
        {\includegraphics[width=0.80in]{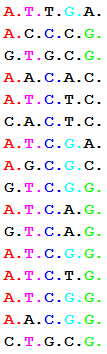}}
        \par\vspace{0pt}
      \end{minipage}%
 \caption{The congruence derivative matrix of periodicity 5 of an artificial DNA sequence (left) and the inferred 16 copies of approximate 5 bp repeats (right). The consensus sequence of the repeats inferred from the matrix is ATCGG in color. The DNA sequence of 80 bp is described in the text.}
 \label{fig:test}
 \end{figure}

\subsection{Computing the distribution uniformity (DU) of a DNA sequence}
Identification of repeats and periodicities in DNA sequences is often realized by Fourier transform. However, Fourier power spectrum may not characterize repetitive elements precisely. It does not identify the consensus repetitive pattern, copy number and perfect level. We propose to employ the periodic nucleotide frequencies as a measure of repeats and periodicities. Previously we showed that the nucleotide distribution on periodic positions correlates with the strength of periodicity 3 \citep{yin2005fourier}. We here generalize to use the nucleotide distribution as the signal measure for any periodicities. Since the CD matrix contains the nucleotide frequencies on periodic positions, the variance of the matrix elements can measure the nucleotide distribution. For the CD matrix of periodicity $p$, the summation of total $4p$ elements of the matrix is equal to length $n$ of the DNA sequence, the  mean of the elements of the matrix is $\frac{n}{{4p}}$. To measure the nucleotide distribution and use it to indicate the periodic signal in a DNA sequence, we define the distribution uniformity (DU) of a DNA sequence using the CD matrix (Definition 2.2).   
\begin{definition}
For a DNA sequence of length $n$, let $f_{i,j}$ be an element of the CD matrix of periodicity $p$, the distribution uniformity of periodicity $p$ of the sequence is defined as
\begin{equation}
DU(p) = \sum\limits_{i = 1}^4 {\sum\limits_{j = 1}^p {(f_{i,j}  - \frac{n}
{{4p}})^2 } } 
\end{equation}
\end{definition}
From Definition 2.2., we notice that the distribution uniformity at periodicity $p$ is an intuitive description for the level of unbalance of nucleotide frequencies on periodic positions. It depends on the quadratic function of the nucleotide frequencies, sequence and periodicity length. Additionally, Definition 2.2 can be rewritten as Equation (3). It shows that for fixed length and periodicity, the distribution uniformity only depends on the squares of nucleotide frequencies. The proof of Equation (3) is provided in the supplementary materials.
\begin{equation}
DU(p) = \sum\limits_{i = 1}^4 {\sum\limits_{j = 1}^p {f_{i,j}^2  - \frac{{n^2 }}
{{4p}}} } 
\end{equation}

For a DNA sequence of length $n$ consisting of tandem perfect repeats of size $p$, each column of the CD matrix contains three zeros and one non-zero element $n/p$. Therefore, the distribution uniformity of tandem perfect repeats is described in Equation (4).

\begin{equation}
\begin{gathered}
  DU_X (p) = \sum\limits_{i = 1}^4 {\sum\limits_{j = 1}^p {(f_{i,j}  - \frac{n}
{{4p}})^2 } }  \hfill \\
   = p\left( {\left( {\frac{n}
{p} - \frac{n}
{{4p}}} \right)^2  + 3\left( {0 - \frac{n}
{{4p}}} \right)^2 } \right) \hfill \\
   = \frac{{3n^2 }}
{{4p}} \hfill \\ 
\end{gathered} 
\end{equation}

Definition 2.2 and Equation (4) indicate that if the length $n$ of a DNA sequence is long or if repeat length $p$ is short, then the nucleotide distribution uniformity becomes large. For a random DNA sequence with uniform nucleotide distribution, each element of the CD matrix of periodicity $p$ is equal to $\frac{n}{{4p}}$, so the $DU(p)$ is zero. Thus the distribution uniformities of any DNA sequences are in range $[0,\frac{{3n^2 }}{{4p}}]$.

A copy number of repeats indicates how many repeats in a DNA sequence or genome. Copy number variations  depends on repeat types and genomes and play an important role in generating variation in population and disease phenotype \citep{mccarroll2007copy}. The strength of a distribution uniformity is impacted by the copy number of repeats. The more copies of repeats, the stronger of the distribution uniformity. We define the copy number of repeats as follows (Definition 2.3).
\begin{definition}
The copy number of the repeats of size $p$ is equal to the division of the sequence length by the periodicity length $p$.
\begin{equation}
CP(p) = \frac{n}
{p}
\end{equation}
\end{definition}

The above definitions indicate that the longer a DNA sequence is, the larger the copy number is and the stronger distribution uniformity is. To consistently compare the strengths of periodic signals in DNA sequences of different lengths, we use the normalized distribution uniformity (NDU), which can be considered as the mean distribution uniformity. It is defined as the distribution uniformity of periodicity $p$ divided by the length of the sequence (Definition 2.4). NDU(p) can be used to indicate for existence of the periodicity $p$ in a DNA sequence.
\begin{definition}
The normalized distribution uniformity (NDU) is equal to the distribution uniformity divided by the sequence length $n$, 
 \begin{equation}
 NDU(p) = \frac{{DU(p)}}
 {n}
 \end{equation}
\end{definition}
From Equation (3), the normalized distribution uniformity of periodicity $p$ of a DNA of length $n$ can be further described as in Equation (7)
 \begin{equation}
  NDU(p) = \frac{1}
  {n}\sum\limits_{i = 1}^4 {\sum\limits_{j = 1}^p {f_{i,j}^2  - \frac{n}
  {{4p}}} } 
  \end{equation}
   
For tandem perfect repeats, we get NDU by Definitions 2.3 and Equation (4) as follows.
\begin{equation}
NDU_X (p) = \frac{{DU_X (p)}}
{n} = \frac{{3n^2 }}
{{4pn}} = \frac{{3n}}
{{4p}} = \frac{3}
{4}CP(p)
\end{equation}

Equation (8) indicates that if the NDU of tandem perfect repeats is larger than 1, the copy number of tandem perfect repeats is at least $4/3$. This fact designates the sequence feature of the NDU threshold value as 1. When the NDU of a DNA sequence is larger than 1, it suggests existence of a significant periodicity and repeats in the sequence. This is the foundation of our algorithm for detecting periodicity and repeats by using the distribution uniformity of nucleotides.
   
\subsection{Identifying the repeat pattern of length $p$ in a DNA sequence}
Using the CD matrix of periodicity $p$, we can determine the dominant nucleotide in each column of the matrix, where the nucleotide has the maximum frequency at the $q$th periodic position $q= 1,2 \ldots,p$ (Definition 2.5). From the positions of dominant nucleotides, we can determine the consensus repeat pattern.

\begin{definition} The dominant nucleotide of the congruence derivative matrix of periodicity $p$ of a DNA sequence is the element with the maximum frequency in each column of the matrix.
\begin{equation}
d_j  = \max (f_{i,j} ),i \in \{ 1,2,3,4\} ,j = 1,2, \ldots ,p
\end{equation}
\end{definition}

The consensus repeat pattern of size $p$ of a DNA sequence may be inferred from the dominant nucleotide frequencies of the congruence derivative matrix of a periodicity $p$. In a CD matrix, $d_j$ is the maximum of the $j$th column, if the corresponding nucleotide in the matrix for $d_j$ is $\alpha  \in \{ G,C,A,T\}$, then nucleotide $\alpha$ appears periodically at the $j$th positions in the sequence. In detail, let $r_j$ be $j$th nucleotide of the repeat unit of size $p$ of the DNA sequence, $j = 1,2, \ldots p$. If $d_j = f_{i,j}$, then $r_j  = \alpha _i,\alpha  \in \{ G,C,A,T\} $. Thus the consensus repeat pattern can be identified.

After the consensus repeat pattern of repetitive elements is determined, a parameter is needed to measure the perfect level of the approximate repeats compared with the tandem perfect repeats of the same length. We define the perfect level $PR(p)$ as the percentage of matched nucleotides in a DNA sequence with tandem perfect repeats of the same length (Definition 2.6).

\begin{definition}
The perfect level $PR(p)$ is the percentage of matched nucleotides in a DNA sequence with tandem perfect repeats of the same length. It is equal to the summation of dominant column elements in the CD matrix divided by the sequence length $n$. 
\begin{equation}
PR(p) = \frac{1}
{n}\sum\limits_{j = 1}^p {d_j } 
\end{equation}
\end{definition}

Since the NDU value of approximate repeats also depends on perfect level, if at least two copies of tandem repeats are required, then from equations (3-10), the perfect level must be at least 2/3 (i.,e., 66.67\%) to make NDU threshold larger than 1. In this study, we use a NDU threshold of 1 (\textit{i.e.}, 100\%) to determine whether a distribution uniformity truly indicates the periodicity.

As an example, from the CD matrix of periodicity 5 in Fig.1, the dominant nucleotide frequencies $d_j$ of the DNA sequence of length 80 bp are [11,11,13,7,10] and the derived consensus repeat pattern is ATCGG. The perfect level computed by Equation (10) is 65\%. It indicates that among repeat elements, 65\% nucleotides match 16 copies of tandem perfect repeats ATCGG.

\subsection{Algorithms}
The inputs for computing the distribution uniformity, copy number, and perfect level are the sequence length $n$ and periodicity length $p$. Given a DNA sequence and periodicity $p$ as inputs, we have the following algorithm, named the distribution uniformity (DU) method, for identifying repetitive elements and periodicities in a DNA sequence.

\begin{algorithm} 
 \SetAlgoLined
 \KwIn{A DNA sequence of length $n$, periodicity $p$}
 \KwOut{DU, NDU, consensu repeat pattern, perfect level, copy number}
 \textbf{Step:}
 initialization \\
 1. Convert DNA sequence into 4D binary indicators.\\
 2. Compute congruence derivative (CD) matrix of periodicity $p$ from the 4D indicators.\\
 3. From the CD matrix, we can compute:\\
 3.1. $DU(p)$ at periodicity $p$ (Equation (3)).\\
 3.2. $NDU(p)$ (Equation (6)).\\
 3.3. Consensus repeat pattern of size $p$ (Equation (9)).\\
 3.4. Perfect level $PR(p)$ (Equation (10)).\\
 3.5. Copy number for the repeat of size $p$ (Equation (5)).\\ 
 
 \caption{The algorithm for identifying repeats and periodicity in a DNA sequence.}
\end{algorithm}

To compute distribution uniformities of different periodicities of a DNA sequence, we first scan the sequence in different periodicity sizes, construct the congruence derivative matrix of each periodicity, and compute the distribution uniformities of these periodicities. The periodicity with the maximum distribution uniformity reflects the dominant pattern of repetitive elements.

Just like the short-time Fourier transform (STFT), to identify the repeat regions in a DNA sequence, we adopt a fixed size window to slide along DNA sequences, and compute the NDU values of different periodicities in the window continuously. The NDU values of periodicities indicate the perfect levels and copy numbers of corresponding repeat regions. 

\section{Results}
\subsection{Periodic analysis of short DNA sequences}
For the artificial DNA sequence as described in Fig.1, the DU method can identify harmonious periodicities 5, 10, 15, 20, 25, etc (Fig.2 (a)). The NDU for periodicity 5 (Fig.2 (b)) and NDU values 1-40 in the sequence using short sliding window of length 20 bp (Fig.2,(c)) conform the repeat structures in the sequence. The results show that the signal of periodicity 5 is clear with very short window size 20 bp. The capacity of capturing periodic signals from short DNA sequences is an advantage of the proposed method.

\begin{figure}[tbp]
          \centering
          \subfloat[]{\includegraphics[width=3.25in]{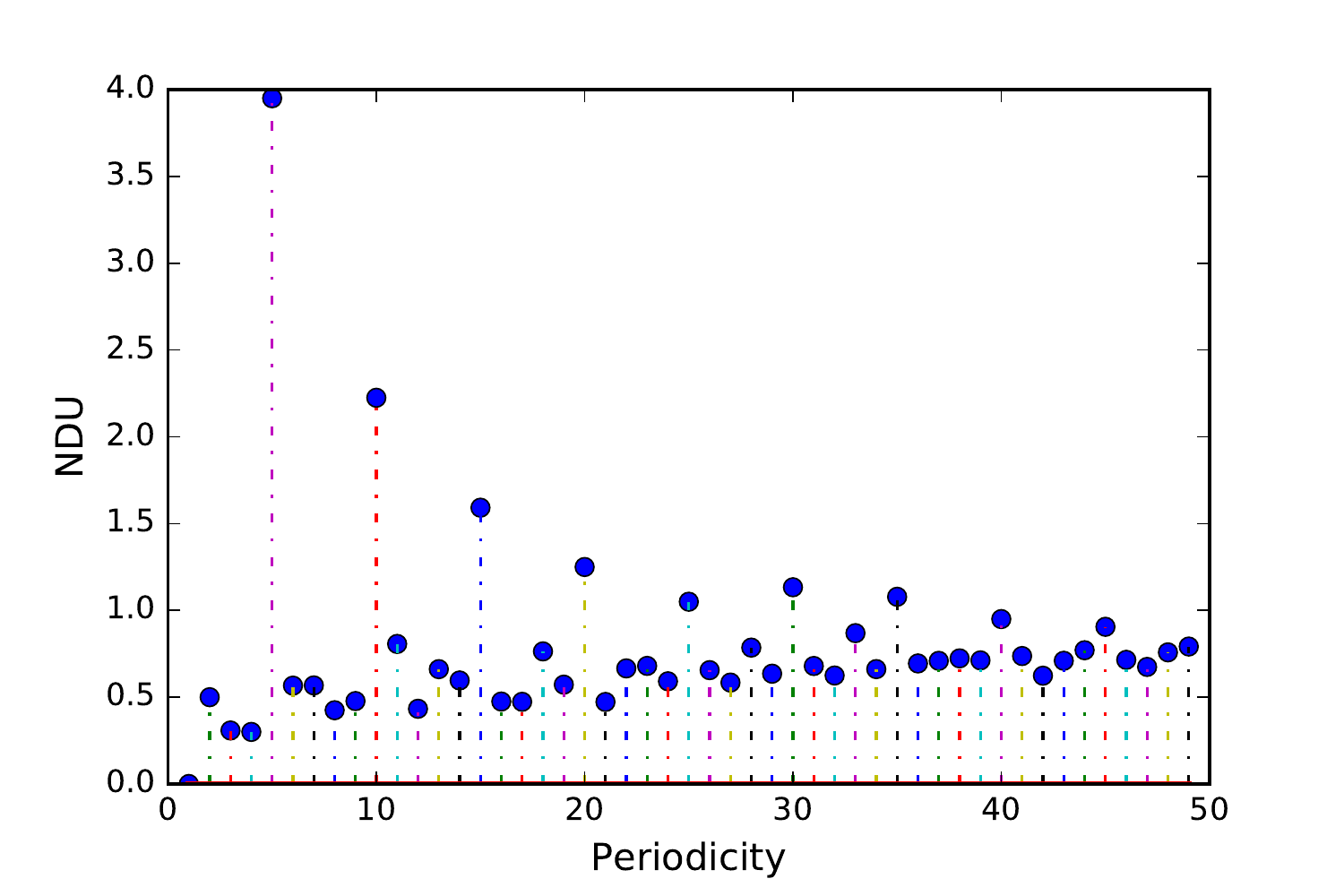}}\quad
          \subfloat[]{\includegraphics[width=3.25in]{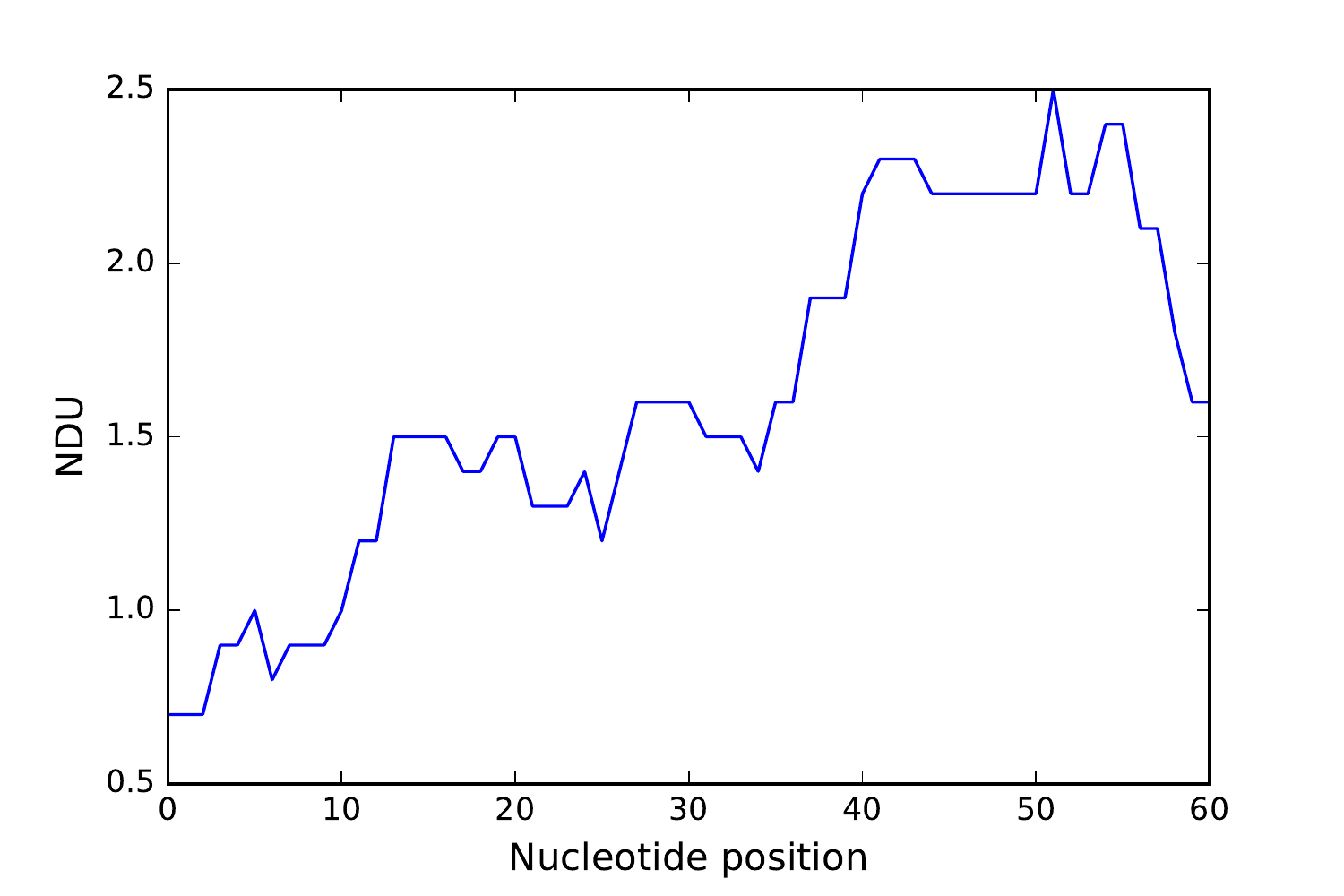}}\quad
          \subfloat[]{\includegraphics[width=3.50in]{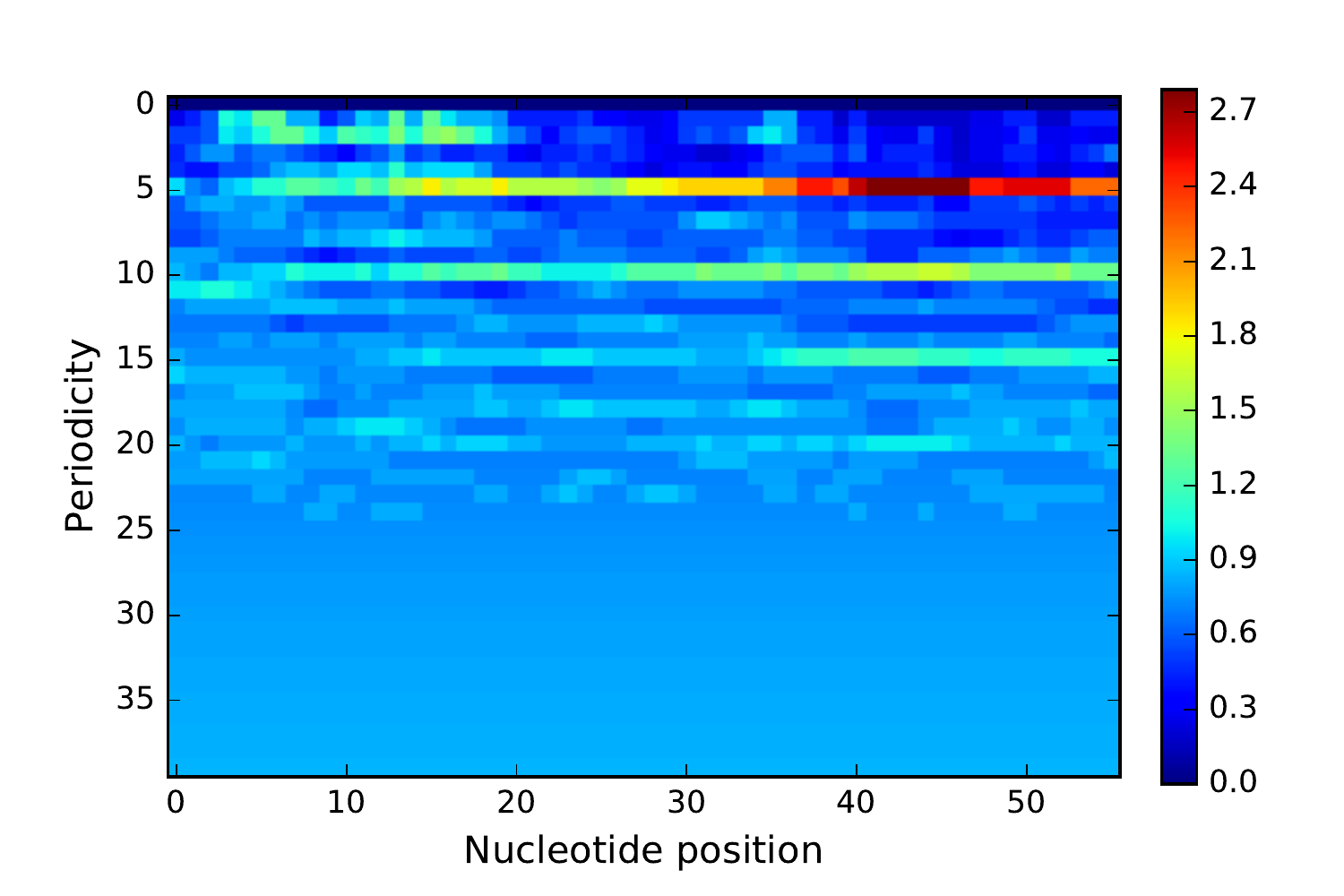}}\quad
          \caption{The distribution uniformity of the artificial DNA sequence. The sequence and its CD matrix are shown in Fig.1. (a) NDU of different periodicities. (b) NDU of periodicity 5 of the DNA sequence by sliding window of 20 bp. (c) NDU of different periodicities in the DNA sequence by sliding window of 20 bp.}
          \label{fig:sub1}
 \end{figure}
\subsection{Periodic analysis of highly complex repeats in DNA sequences}
The effectiveness of the proposed algorithm is tested on \textit{Homo sapiens} collagen type IV alpha 6 chain (COL4A6) (GenBankID:NC$\_$001847, 6618 bp). This gene contains repeats of length 9 bp by the EPSD method \citep{gupta2007novel}. The result by the DU method shows that the sequence contains repeats of 3 bp, 6 bp, 9 bp (Fig.3(a)). The sliding window approach shows the location and strength of the periodicity 9 (Fig.3(b)) and periodicities $1~100$ (Fig.3(c)). The corresponding periodicity strength, NDU, perfect level, consensus pattern, and copy number can be revealed by the method (Table 1.). Table 1 shows that the perfect levels of periodicities 3 , 6 and 9 are similar, but the periodicity 3 has the largest mean NDU because of the highest copy number of periodicity 3. The periodicities in this DNA sequence identified by the proposed method are consistent with the EPSD method \citep{gupta2007novel}. The results demonstrate the effectiveness of the proposed method in capturing different repeats and periodicities.

\begin{figure}[tbp]
         \centering
         \subfloat[]{\includegraphics[width=3.25in]{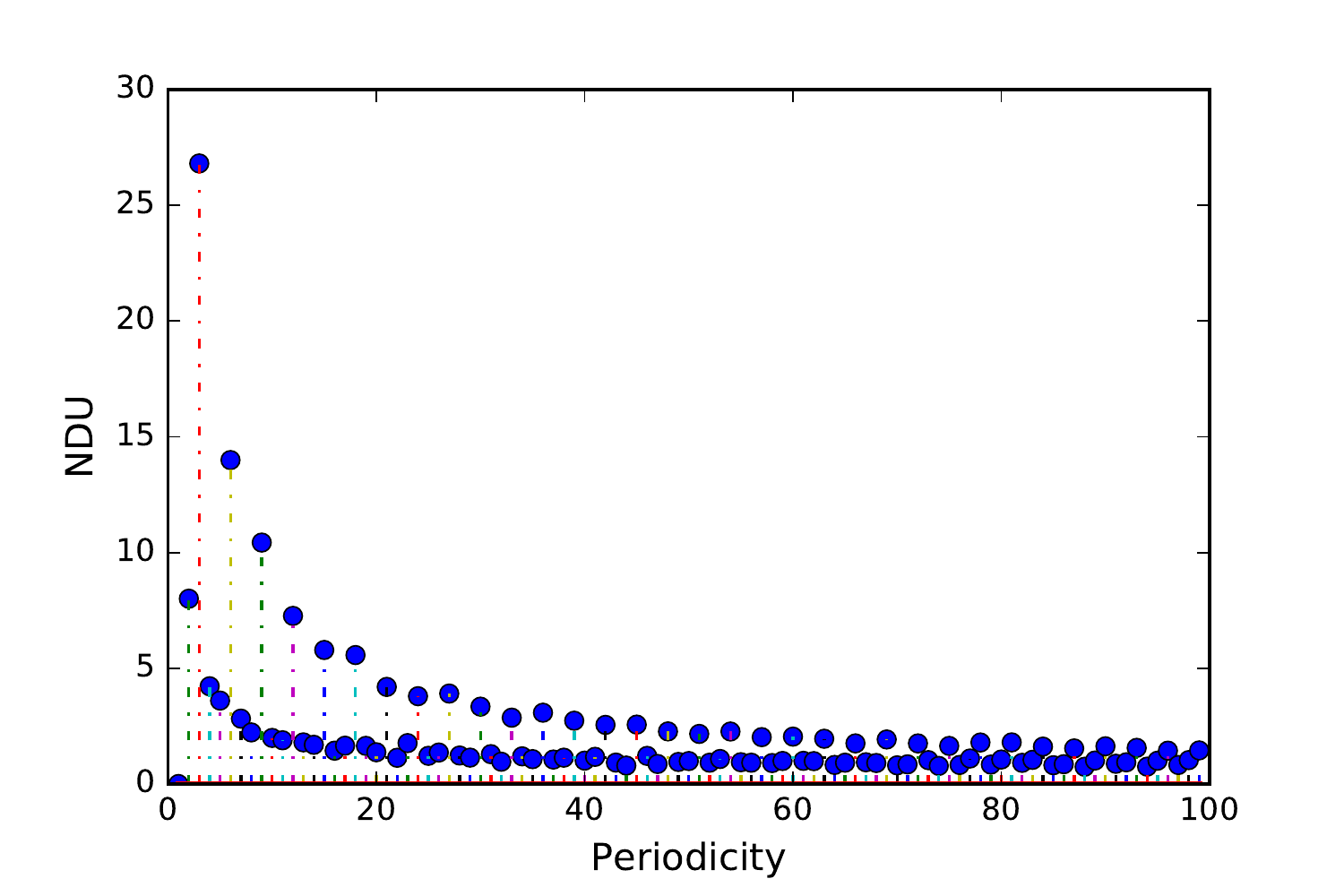}}\quad
         \subfloat[]{\includegraphics[width=3.25in]{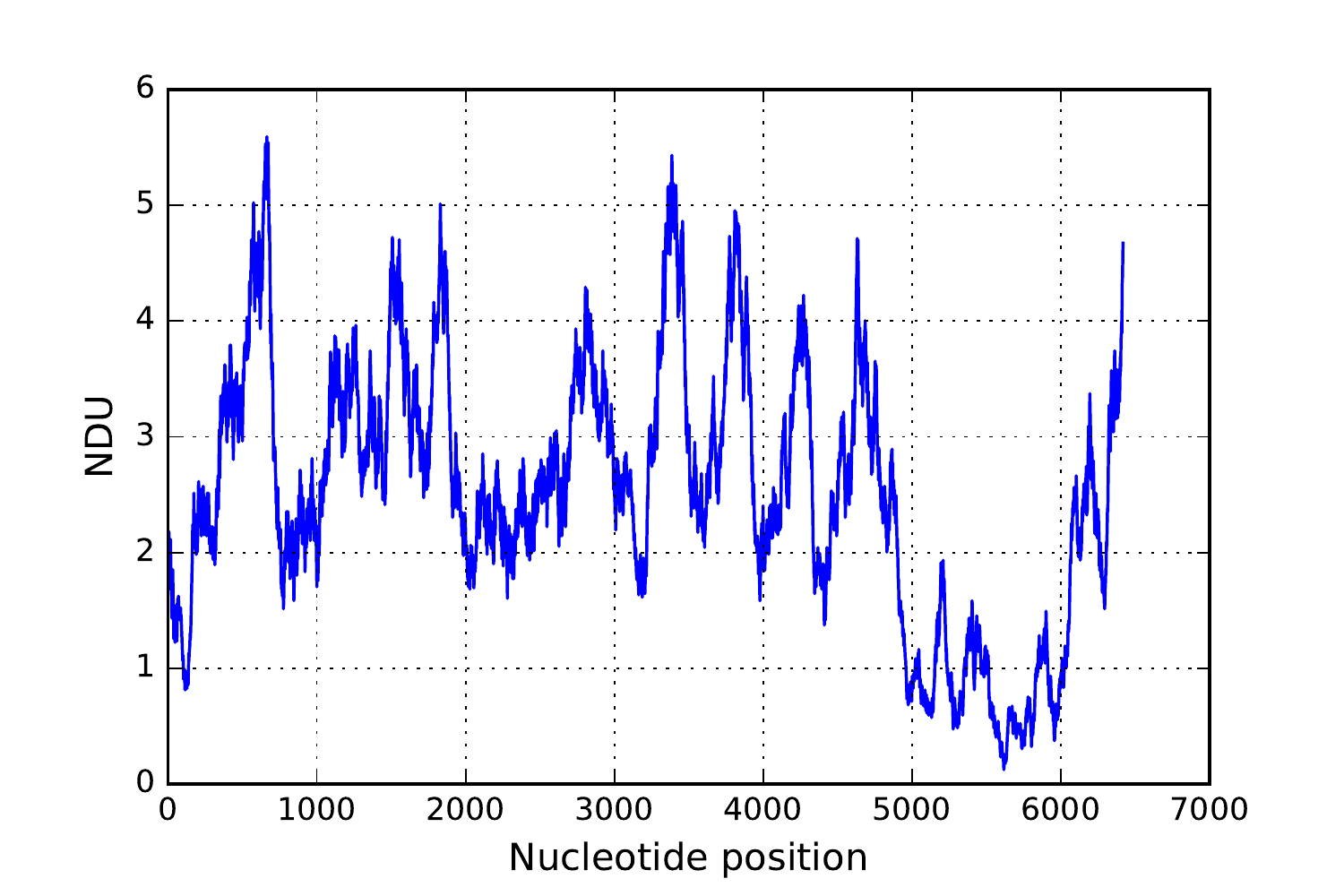}}\quad
         \subfloat[]{\includegraphics[width=3.50in]{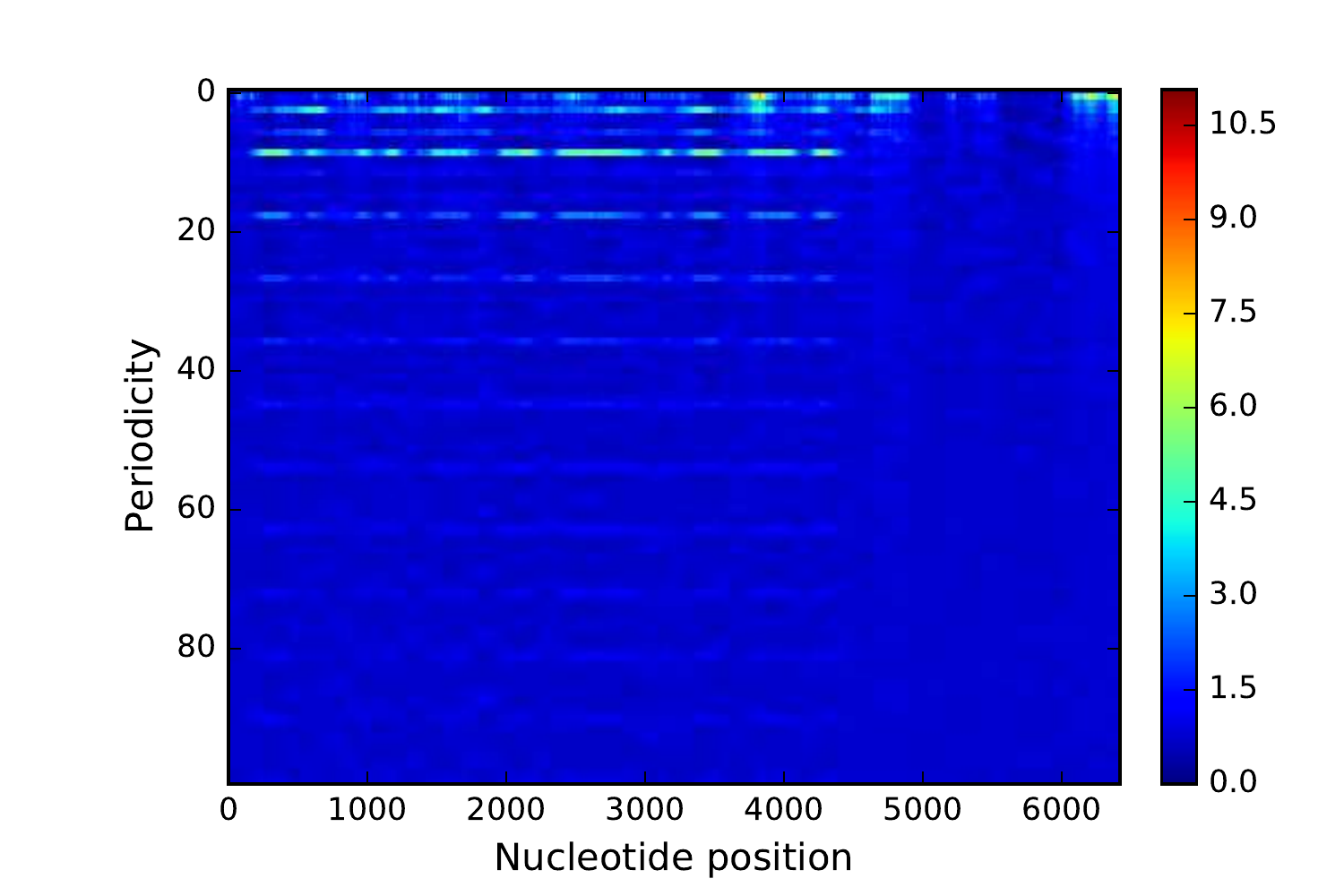}}\quad
         \caption{The distribution uniformity analysis of Homo sapiens collagen type IV alpha 6 chain (COL4A6) (GenBankID:NC$\_$001847). (a) NDU of different periodicities. (b) NDU of periodicity 9 of the DNA sequence the sliding window approach. (c) NDU of different periodicities in the DNA sequence by sliding window approach. The window size is 200 bp.}
         \label{fig:sub1}
\end{figure}

  \begin{table}[ht]
   \caption{Characterization of the repeats in human COL4A6 gene (GenBankID:NC$\_$001847).}
   \centering 
   \begin{tabular}{l*{5}{l}r}
   \hline\hline
   p  & NDU(p) & PR(p) & pattern & copies \\
   \hline
    3 & 26.81036 & 0.3262  & GGA       & 2206\\
    6 & 13.9946  & 0.3262  & GGAGGA    & 1103 \\
    9 & 10.4310  & 0.3312  & GCAGGAGGT & 735 \\
   \hline\hline
   \end{tabular}
   \label{table:nonlin} 
   \end{table}
   
To test if the DU method can identify complex repeats, we use the method to analyze Human microsatellite repeats (GenBank locus:HSVDJSAT, 1985 bp). The DNA sequence contains variable length tandem repeats (VLTRs) \citep{hauth2002beyond}. The DU method may detect accurately all different the short periodicities and corresponding perfect levels and copy number of the repeats in the full DNA sequence (Fig.4(a), Table 2). As an example, the sliding window approach for the NDU of periodicity 7 indicates that the 7 bp repeats are between positions 1000 and 1550 when using the NDU threshold as 1 (Fig.4(b)). The method may locate the positions of other repeats from the corresponding periodicities (Fig.4(c)). Specially, in the region between 1000 and 1550 bp, two long repeats of 18 bp and 19 bp can be identified using the DU method (Table 3). These results agree with the previous studies by statistical methods \citep{hauth2002beyond,gupta2007novel}, however, the previous studies need different threshold settings and sometimes get conflicting results.
\begin{figure}[tbp]
         \centering
         \subfloat[]{\includegraphics[width=3.25in]{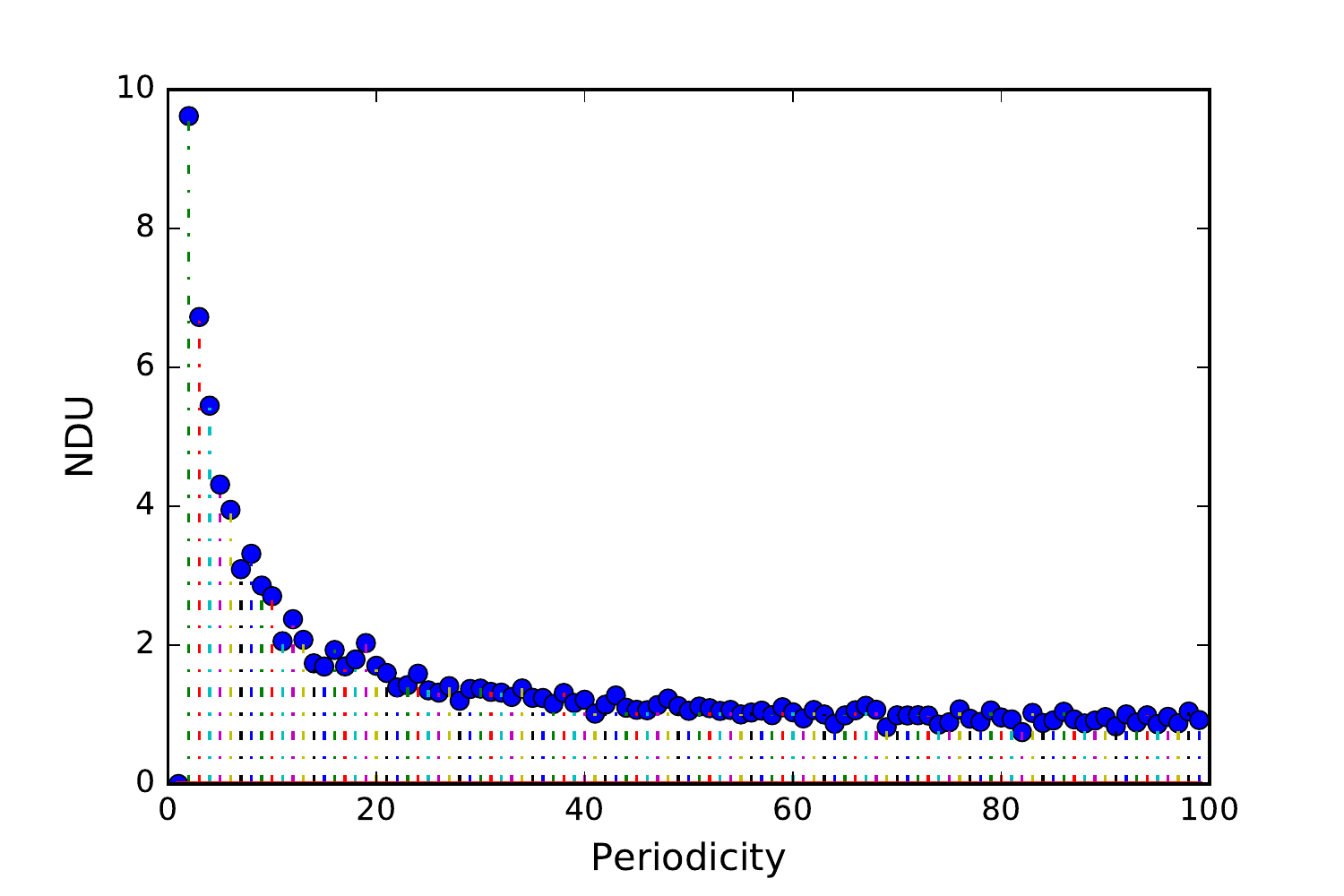}}\quad
         \subfloat[]{\includegraphics[width=3.25in]{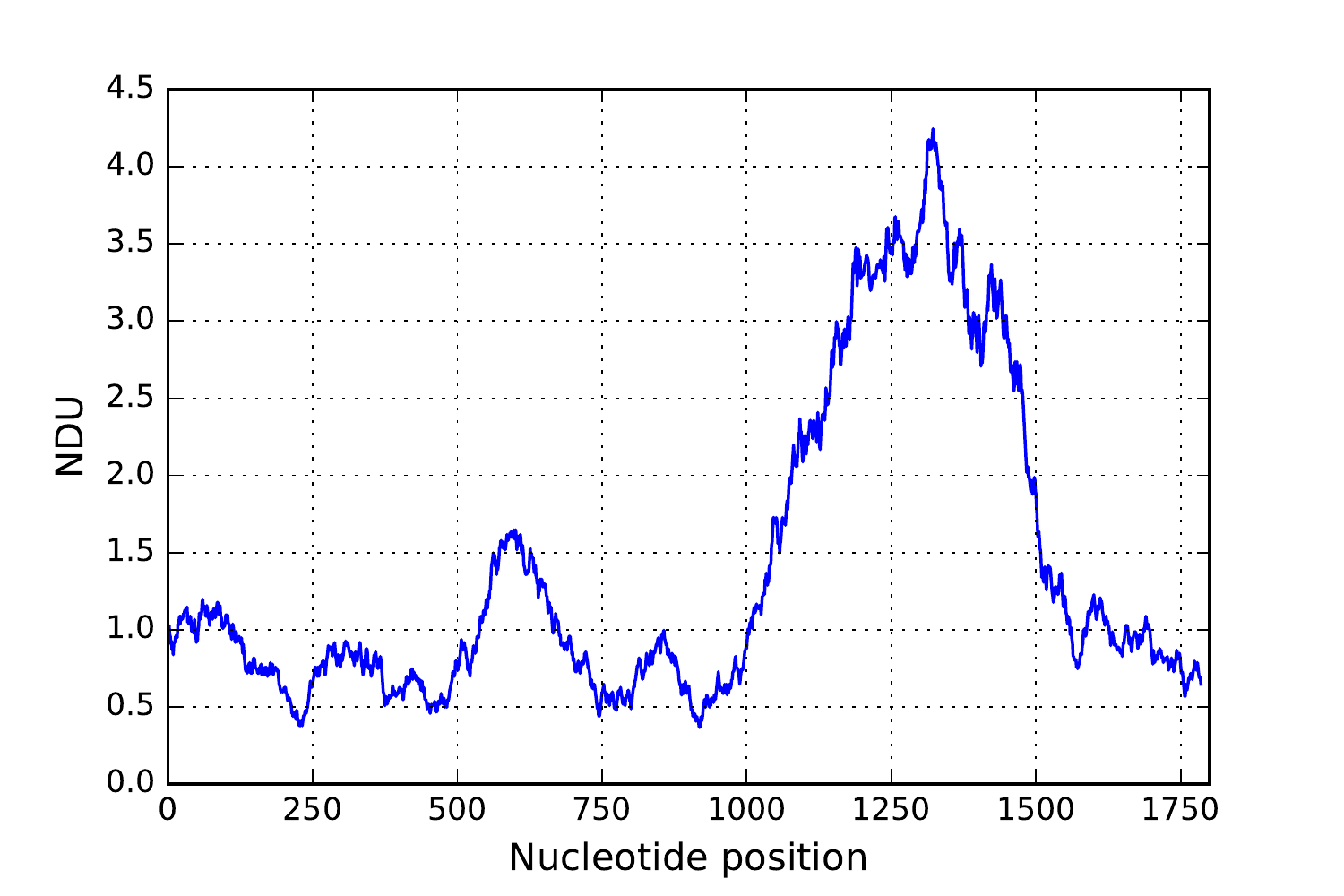}}\quad
         \subfloat[]{\includegraphics[width=3.50in]{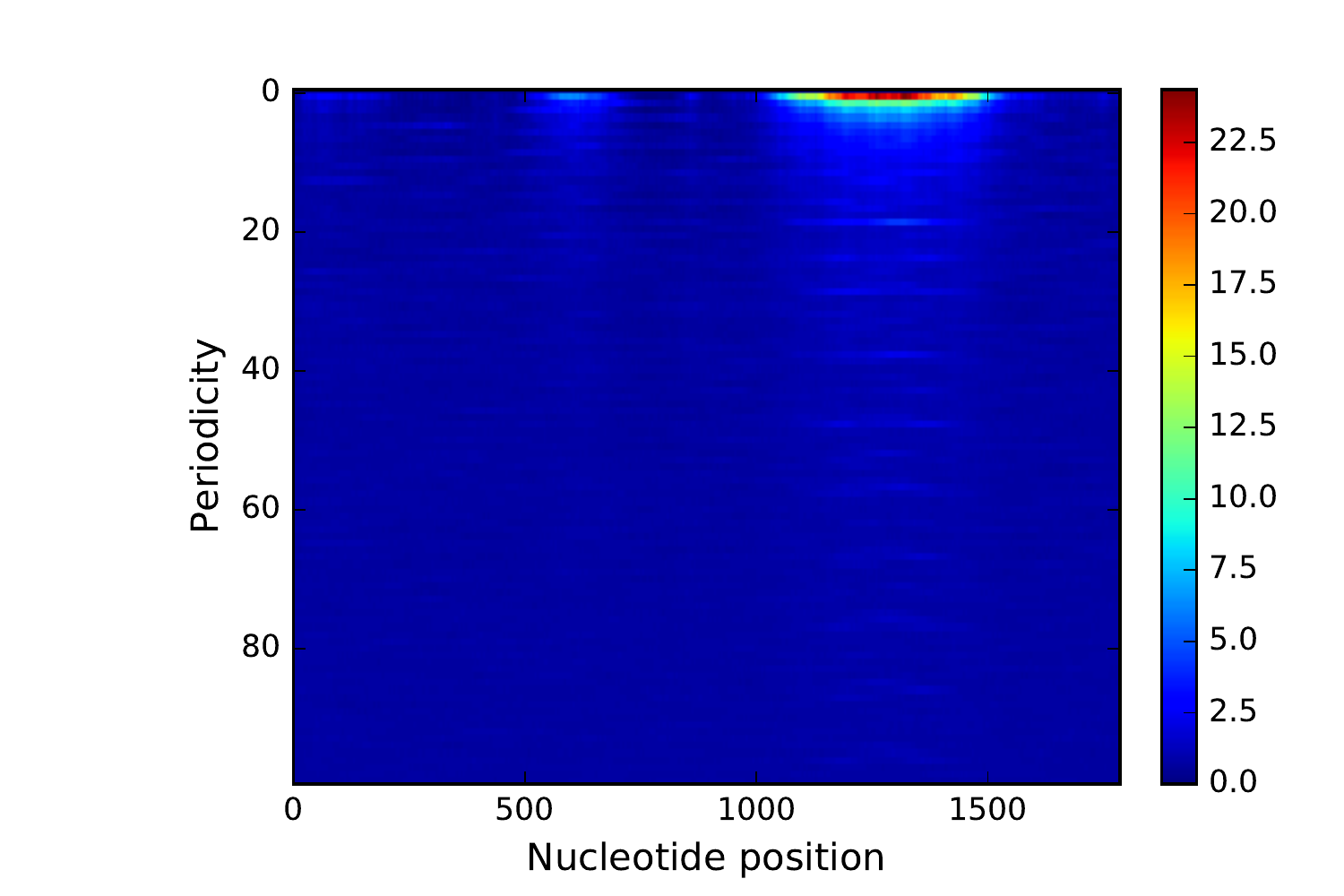}}\quad
         \caption{The distribution uniformity analysis of variable length tandem repeats (VLTRs). (a) NDU of different periodicities. (b) NDU of periodicity 5 of the DNA sequence by the sliding window approach. (c) NDU of different periodicities of the DNA sequence by the sliding window approach. The window size is 200 bp.}
         \label{fig:sub1}
\end{figure}

 \begin{table}[ht]
   \caption{Characterization of repeats in the Human microsatellite sequence (GenBank locus:HSVDJSAT, region 1-1985 bp).}
   \centering 
   \begin{tabular}{l*{5}{l}r}
   \hline\hline
   p  & NDU(p) & PR(p) & pattern & copies \\
   \hline
    7 & 3.0935 &  0.3274  & GGGGGGG       & 283\\
    8 & 3.3153  & 0.3314  & GGAGGGTG    & 248 \\
    9 & 2.8575  & 0.3312  & GGGGGGGGA &  220 \\
    10 & 2.7049  & 0.3314  & GGAGTGGGGG &  198 \\
    16 & 1.9299  & 0.3324 & GGAAGGTGGGAGGGTG &  124 \\
    17 & 1.6939  &  0.3324 & GGAAGGTGGGAGGGTG &  116 \\
    18 & 1.7932  &  0.3340 &  AGGGGGAGGGGGGAGGGA &  110 \\
    19 & 2.0291  &  0.3420 &  GGCGGGGGTAGGCGGGGAG &  104 \\
   \hline\hline
   \end{tabular}
   \label{table:nonlin} 
   \end{table}

\begin{table}[ht]
   \caption{Characterization of long repeats in the Human microsatellite sequence (GenBank locus:HSVDJSAT, region 1000-1550 bp).}
   \centering 
   \begin{tabular}{l*{5}{l}r}
   \hline\hline
   p  & NDU(p) & PR(p) & pattern & copies \\
   \hline
    18 & 1.9875  &  0.4560 &  GGGGGGGGGGGGGGGGGG &  27 \\
    19 & 3.5210  &  0.5200 &  CTGGGAGGGCTGGGAAAGG &  26 \\
   \hline\hline
   \end{tabular}
   \label{table:nonlin} 
   \end{table}
   
\subsection{Computational complexity}
The computation complexity of the DU method for specific periodicity p of DNA sequence of length $n$ only involves computing CD matrix ${\rm O}(n)$ and small number of multiplication of entries in the matrix ${\rm O}(p^2+p))$. Because periodicity p is much smaller than n, ${\rm O}(p^2 +p))$ is very small, the computation complexity of the DU method is approximate linear with the sequence length, ${\rm O}(n)$. In contrast, using Fourier transform in detecting repetitive elements has high computational complexity. The fast Fourier transform (FFT) needs ${\rm O}(n\log n)$ computational time. Because only distribution uniformities at specific periodicities are needed for detecting repetitive elements of interested lengths, instead of DUs for all periodicities, the performance of the DU method for specific periodicities is more efficient than Fourier transform. Performance tests were performed on a PC with configuration as Intel Core i5 processor, 6G RAM. The DU method has linear runtime for any sequence lengths as shown in Fig.5. This result verifies the complexity analysis, showing high efficiency of the DU method.
\begin{figure}[tbp]
            \centering
            \subfloat[]{\includegraphics[width=3.5in]{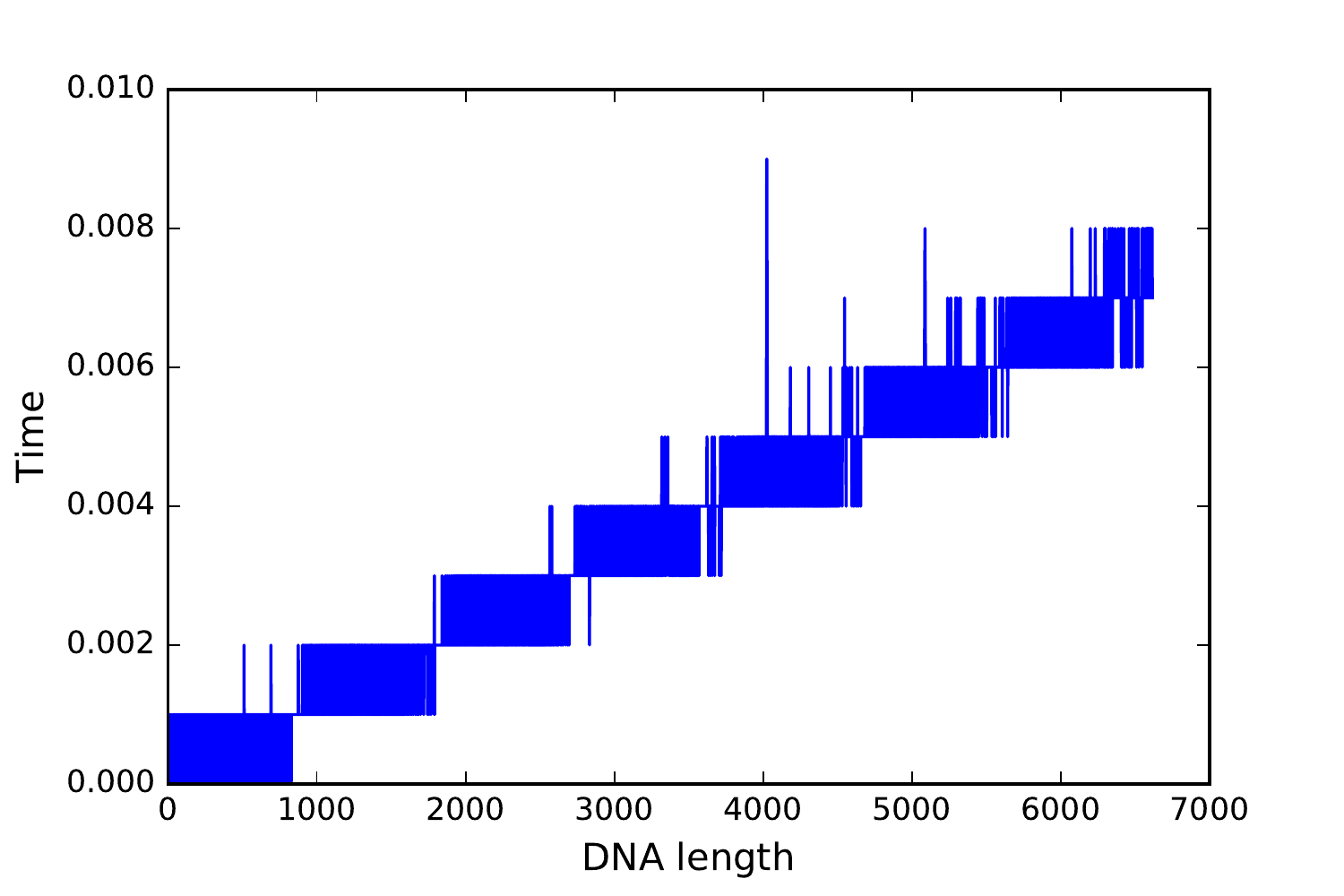}}\quad
            \caption{Running time of the DU method on DNA sequences of different lengths.}
            \label{fig:sub1}
\end{figure}
  
\section{Discussion}  
Using the distribution uniformity of nucleotides in DNA sequences, we establish an \textit{ab initio} method in identifying repeats and periodicities in the DNA sequences. The distribution uniformity intuitively depicts the level of unbalance of nucleotide frequencies on periodic positions. The unbalance  of nucleotide frequencies determine the nature of repeats and periodicities in DNA sequences. The method does not rely upon previously known repeat sequence information and is natural, effective and convenient. It can quantitatively describe a DNA repeat with NDU, consensus repeat pattern, perfect level, and copy number. 

This study shows that the periodicities of genome is contributed by the tandem repeats of the corresponding length. For example, a strong periodicity of length 5 is caused by large copies of 5 bp repeats. The approximate repeats render high frequencies of nucleotides at periodic positions in DNA sequences. The consensus repeat patterns  identified by the proposed method can be considered as the original sequences from which the approximate repeats have been evolved in evolutionary history. The perfect level of the approximate repeats represent the evolutionary progress. Thus the repetitive sequences can be used in phylogenetic analysis \citep{versalovic1991distribution}.

Unlike the Fourier transform method, the DU method can reveal long harmonious periodicities, which are often missed in Fourier transform. The DU method also directly computes the distribution uniformity for interested periodicities, whereas Fourier transform computes the power spectrum for frequencies, and there is no a straightforward solution to convert it into the power spectrum for periodicities. 

Despite the efficacy in capturing repeats and the corresponding features by the proposed method, there are some limitations that the DU method cannot solve. One limitation is that it can not identify interspersed repetitive elements or repeats interrupted by deletion mutations. We will address this limitation in our future research.

The frequencies of nucleotides vary significantly across eukaryotic genes and may present specific regulatory motifs \citep{louie2003nucleotide}. Because the nucleotide distribution represents the essential characteristics of a DNA sequence, the distribution uniformities of different periodicities may have broad applications in functional analysis of genomes.  
 
%
%
\section*{acknowledgement}
We appreciate Professor Jiasong Wang at Department of Mathematics, Nanjing University, China, for his mentorship and valuable advice for this research. We are grateful to Emily Yin at Indiana University and Tung Hoang for proof reading of the paper.
%
%

\bibliographystyle{elsarticle-harv}
\bibliography{../References/myRefs}

\section*{Supplementary Materials} 
\begin{theorem}
The distribution uniformity of a periodicity $DU(p)$ is determined by the nucleotide frequencies at $p-$th periodic positions $f_{i,j}, i = 1,2,3,4,j = 1,2 \ldots ,p$, DNA length $n$ and periodicity length $p$. 
$$
DU(p) = \sum\limits_{i = 1}^4 {\sum\limits_{j = 1}^p {f_{i,j}^2  - \frac{{n^2 }}
{{4p}}} } 
$$
\end{theorem}
Proof:
$$
\begin{gathered}
  DU(p) = \sum\limits_{i = 1}^4 {\sum\limits_{j = 1}^p {(f_{i,j}  - \frac{n}
{{4p}})^2 } }  \cdot  \hfill \\
   = \sum\limits_{i = 1}^4 {\sum\limits_{j = 1}^p {(f_{i,j}^2  - \frac{{nf_{i,j} }}
{{2p}} + \frac{{n^2 }}
{{16p^2 }})} }  \hfill \\
   = \sum\limits_{i = 1}^4 {\sum\limits_{j = 1}^p {f_{i,j}^2  - \frac{n}
{{2p}}\sum\limits_{i = 1}^4 {\sum\limits_{j = 1}^p {f_{i,j} } }  + \sum\limits_{i = 1}^4 {\sum\limits_{j = 1}^p {\frac{{n^2 }}
{{16p^2 }}} } } }  \hfill \\
   = \sum\limits_{i = 1}^4 {\sum\limits_{j = 1}^p {f_{i,j}^2  - \frac{n}
{{2p}} \cdot n + } } \frac{{n^2 }}
{{16p^2 }} \cdot 4p \hfill \\
   = \sum\limits_{i = 1}^4 {\sum\limits_{j = 1}^p {f_{i,j}^2  - \frac{{n^2 }}
{{2p}} + } } \frac{{n^2 }}
{{16p^2 }} \cdot 4p \hfill \\
   = \sum\limits_{i = 1}^4 {\sum\limits_{j = 1}^p {f_{i,j}^2  - \frac{{n^2 }}
{{4p}}} }  \hfill \\ 
\end{gathered} 
$$
\end{document}